\documentstyle[aps]{revtex} 
\begin{document}
\begin{center}

{\large \bf LOCALIZATION PROPERTIES OF THE PERIODIC 
RANDOM ANDERSON MODEL}\\
\vskip .5cm
{\bf M. Hilke \footnote{ Present Address: Dpt. of electrical engineering, 
Princeton University, NJ-08544, USA.} }\\
\vskip.2cm
{\small Universit\'e de Gen\`eve,  D\'epartement de Physique Th\'eorique\\
          24 Quai Ernest-Ansermet CH-1211 Gen\`eve 4\\ Switzerland}\\
\vspace*{1cm}

\parbox{10cm}{
\baselineskip=14pt
 We consider   diagonal disordered one-dimensional Anderson models with an  
underlying periodicity. We assume the simplest periodicity, i.e., we 
have essentially two lattices, one that is composed of the random potentials 
and the other of non-random potentials. Due to the periodicity 
special resonance energies appear, which are  related to the 
lattice constant of the non-random lattice. Further on two different 
types of behaviors are observed at the resonance energies. When a random 
site is surrounded by non-random sites, this model 
exhibits {\em extended} states at the resonance energies, whereas otherwise
all states are localized with, however,  an increase of the localization 
length at these resonance energies. We study these resonance energies 
and evaluate the localization length and the density of states around these 
energies.

}
\end{center}

\vskip1cm

Localization properties of disordered systems were first examined in 
tight-binding models by 
Anderson \cite{anderson}, who 
 showed that certain states are localized due to disorder.  
 His result 
was generalized by Mott and Twose \cite{mott} and Landauer 
\cite{landauer} who conjectured 
and later several authors \cite{abou,kunz}
proved that, in one 
dimension,  all 
states 
are localized for any amount of disorder.
 
Recently, however, there have been a number of experiments in 
quasi-one dimensional systems which exhibit unusual high conductivities.
These systems are polymers as well as mesoscopic rings 
\cite{dunlap2,ishiguro,advances}. It seems therefore of great importance to 
study delocalization mechanism in disordered systems. As has 
already been pointed out in some recent works on disordered systems, 
correlations in disorder can be a driving force for delocalization 
in one dimension 
\cite{dunlap2,flores,dunlap,bovier,hilke,evan,sanchez} and in two 
dimensions \cite{hilke2}.
The new approach in this paper is to consider systematically the effect of 
a deterministic periodic potential, as source of correlations in the disorder.

In this study we  consider tight-binding models related to the 
original Anderson model. The periodicity is introduced by considering 
two underlying lattices, of which one is composed by the random sites 
and the other by the deterministic sites. In addition we suppose that 
all deterministic sites are constant. This article is  divided 
in two   parts. In the first part we consider the special case 
where each random site is surrounded by at least one 
  constant neighbor site. In this case as discussed below there exist 
discrete resonance energies for which the states are {\em 
overall extended}, i.e., with an infinite localization length. 
In the second part, where the restriction 
above does not apply, we find the same resonance energies. The only 
{\em essential } difference is that instead of having an infinite 
localization length these states  present only 
 an enhancement of this length at 
these critical energies.

In the usual diagonal disordered Anderson model with uncorrelated disorder
on each
site the localization length can be evaluated and yields
$L_c(E)=24(4-E^2)/W^2$
\cite{thouless2}, where $W$ is the width of  the disorder potential
 distribution, for small
disorder.
The consequence is that all states are localized for this model. This is,
however,
only true if we take the average over different configurations of impurities,
as
otherwise it is well known that so called Azbel \cite{azbel} resonances can
appear
for a given
configuration. These resonances of extended states, however, disappear when
we average
over different configurations.

For the first part we start with the 
following Anderson model                                 
\begin{equation}
(V_l-\epsilon)\Psi_l+\Psi_{l+1}+\Psi_{l-1}=0
\end{equation}
where 
$V_l$ is non-zero only if $l$ is a multiple of $d$, where $d$ is an 
integer, i.e.,  
 $V_{dl}$ are random and $V_{dl+1}=\cdots=V_{d(l+1)-1}=0$. The case 
where the deterministic sites are non-zero but constant, is trivially 
obtained by shifting the energy.

The method used to solve this model was developed by Erd\"os and Herndon 
\cite{erdos} and 
later simplified by Felderhof \cite{felderhof}. The idea is 
the following: We suppose that between impurities the solution can be written as the sum of an 
incident plane wave and a reflected plane wave, i.e.,
\begin{equation}
\Psi_l=A_ne^{ikl}+B_ne^{-ikl} \quad,\quad X_{n-1}<l<X_n,
\end{equation}
where $X_n$ are the positions of the impurities with value $V_{dn}$. 
Inserting this in (1) 
yields the following transfer matrix relation for $d>1$:
\begin{equation}
\left(\begin{array}{c}A_{n+1}\\B_{n+1}\end{array}\right)=\left(\begin{array}{cc}\alpha_n & 
e^{-2ikX_n}(-iW_n)\\e^{2ikX_n}(iW_n) & \alpha_n^*\end{array}\right) 
\left(\begin{array}{c}A_{n}\\B_{n}\end{array}\right),
\end{equation}
where $W_n=V_{dn}/2\sin k $, $\alpha_n=1+iW_n $ and $2\cos k=\epsilon $. Instead of considering this 
transfer matrix Felderhof uses the 3-vector transfer matrix namely 
\begin{equation}
\Gamma_n=\left(\begin{array}{ccc} 1-W_n^2-2iW_n & -W_n^2-iW_n & -W_n^2\\
2W_n^2+2iW_n & 1+2 W_n^2 & 2 W_n^2-2iW_n\\-W_n^2 & -W_n^2+iW_n & 1-W_n^2+2 i W_n\end{array}\right)
\end{equation}
\begin{equation}
\mbox{and}\quad G_n=\left(\begin{array}{ccc} e^{2ik(X_n-X_{n-1})} & 0 & 0\\
0 & 1& \\ 0 & 0 & e^{-2ik(X_n-X_{n-1})}\end{array}\right).                                
\end{equation}
The main result of Felderhof is to obtain  $R/T=(P(2,2)-1)/2$, where $R$ is the
reflection coefficient and $T$ the transmission coefficient and 
\begin{equation}
P=\Gamma_n\cdot G_n\cdot\Gamma_{n-1}\cdot G_{n-1}\cdots
\end{equation} 
There are essentially two cases which can be solved analytically in this approach. 
The first one which was studied by Felderhof, who considered the limit 
$k\rightarrow\infty$ and found that the average resistance grows exponentially 
with the number of scatterers or Anderson localization. In our case we only consider the band center,
 i.e., where $E=0$ and when  $d$ is even we have, as 
$E=2 \cos k$ and as $X_n-X_{n-1}=d$, 
that $G_n=I$, where 
$I$ is the identity matrix. Calculating (6) yields the surprisingly simple expression 
\begin{equation}
P(2,2)=2\left(\sum_{n=1}^N W_n\right)^2+1.    
\end{equation}
This shows that when the sum of the impurities is zero the reflection coefficient vanishes. This 
result can be extended to the case where $w=\sum_{n=1}^N W_n$ is non-zero by redefining 
$E=\epsilon-w/N=2\cos k$, which
ensures that $\sum_n (W_n-w/N)=0$ and implies that the reflection coefficient vanishes when 
$\epsilon=w/N$. This result states that we have {\em total transmission} for this model. In fact 
if we suppose that the average of $V_n=0$ then $w/N\rightarrow 0$ for $N\rightarrow\infty$, 
due to central limit theorem. This therefore implies that we have total transmission at  the 
band center in the thermodynamic limit.

Above we showed that we get total transmission at the critical energy. It is straightforward to see 
that the state at $\epsilon=0$ is {\em 
overall extended}, 
as one only needs to suppose that $\Psi_{dl}=0$ and one is left 
with an ordered 
Anderson model.
In the following we study 
the dependence on energy of the localization length around the critical energy. Starting again
from equation (1) and for $d$ even, we renormalize this equation as follows

\begin{equation}
(W_{2l+1}-\epsilon)\Psi_{2l-2}+\Omega_{2l}(\epsilon)\Psi_{2l}+(W_{2l-1}-
\epsilon)\Psi_{2l+2}=0,
\end{equation}
where 
\begin{equation}
\Omega_{2l}(\epsilon)=W_{2l+1}+W_{2l-1}-2\epsilon-(W_{2l+1}-
\epsilon)(W_{2l}-\epsilon)
(W_{2l-1}-\epsilon).
\end{equation}
 Furthermore, in our diluted model $W_{2l+1}=0$, 
which when 
inserted in (9), yields
\begin{equation}
\Psi_{2l-2}+\left(2+\epsilon (W_{2l}-\epsilon )\right)\Psi_{2l}+
\Psi_{2l+2}=0,
\end{equation}
for $\epsilon\neq 0$. This last model was extensively studied in the 
limit $\epsilon\ll1$ by 
Derrida and Gardner \cite{derrida}. They calculated the 
complex Lyapounov exponent $\gamma$, where the real part corresponds to the 
inverse localization 
length and the imaginary part to $\pi$ times the integrated density of 
states. Their results 
can be expressed as follows:
\begin{equation}
\left\{\begin{array}{l}
Re(\gamma) \simeq K_1\epsilon^{2/3}\langle W^2\rangle ^{1/3}\nonumber\\
Im(\gamma) \simeq K_2\epsilon^{2/3}\langle W^2\rangle ^{1/3},
\end{array}\right. 
\end{equation} 
where $K_1=0.29\dots $ and $K_2=0.16\dots $ and $\langle\cdot\rangle$ is 
the average over all 
impurities.
From (11) it is straightforward that the inverse localization length $L_c^{-1}$ scales as 
\begin{equation}
L_c^{-1}\sim\epsilon^{2/3}\langle W^2\rangle ^{1/3}
\end{equation} 
and the density of states is 
\begin{equation}
\rho(\epsilon)=\partial_{\epsilon}Im\gamma(\epsilon)\sim\epsilon^{-1/3}.
\end{equation}
Above we only considered the extended states at $\epsilon=0$ for $d$ even, but there exist 
in general $d-1$ energies at which the states are extended. For $d=3$ for example we have 
delocalized states for $\epsilon=-1$ and $\epsilon=1$. 
For any $d$ they can be easily evaluated as they are the 
roots of $(1,0)\cdot T_{d}\cdot{\left(1\atop 0\right)}=0$, where  
\begin{equation}
T_d=\prod_{n=1}^{d-1}\left({\epsilon \atop 1} {-1\atop 0}\right).
\end{equation}
The solutions can be written as $\epsilon=2\cos{n\pi/d}$, where $n$ is an integer with $|n|<d$.

 These critical energies correspond to the 
resonance energies discussed by Derrida and Gardner \cite{derrida}, for which their expansion in the 
low disorder limit is non-trivial. The two uppermost curves in  fig. 1 show
 the localization length 
as a function 
of the energy. One clearly sees the infinite localization length at
the critical energies. 

\input epsf
\begin{figure}
\hskip 2.5cm
\epsfxsize=9cm
\hspace*{1.5cm}\epsfbox{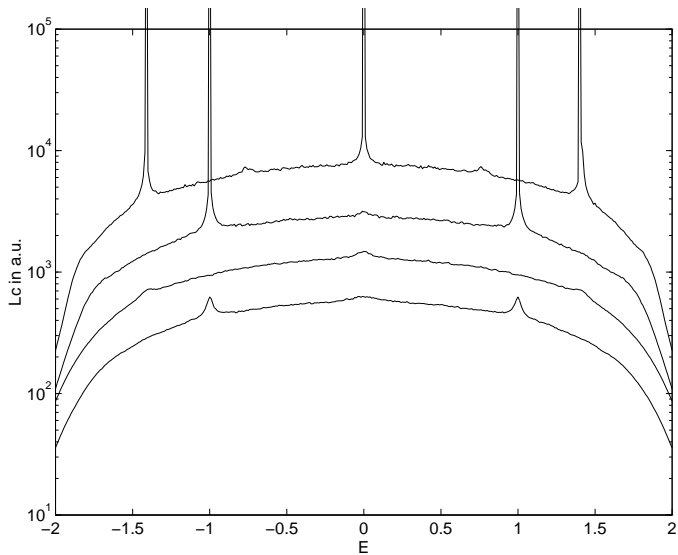}
\vskip .3cm
\caption{Localization length in arbitrary units as a function of  energy. 
The two uppermost curves are from the case discussed in the first part with 
extended states and $d=4$, respectively, $d=3$. The two lowest curves  
represent the case where the localization length is not infinite but enhanced 
at the resonance energies with $d=4$ respectively $d=3$.
}
\end{figure}
\vskip .2cm

In the following we study numerically the case where we can have two 
neighboring random sites. In general one obtains a similar result 
than for the completely random model, where in first order perturbation  
$L_c\sim(4-E^2)$ \cite{thouless2}. The changes occur at the resonance energies discussed above. 
In fact at these energies we have an enhancement of the localization length, as 
shown in fig. 1 for the two lowest curves.

It is interesting to note that the peaks of the localization length correspond exactly 
to the resonance energies discussed above. The relative enhancement however decreases 
with increasing $d$ and in the limit $d\longrightarrow\infty$ we recover the 
usual uncorrelated result. The plot is shown for a system size of $1000$ 
and averaged over a thousand configurations. $d=3$, corresponds to the case 
where every third site is non-random and the sites in between are random. 
This is opposite to the case discussed in the first part, and shown in 
the second uppermost curve of fig. 1, where  every third site is random 
and the sites in between are non-random.

This last study demonstrates that a periodic correlation in the disorder is 
not enough in order to completely delocalize some states. This correlation 
 enhances 
the localization length at some energies related to the periodicity. It 
appears that an important factor is the isolation of the random sites. For 
different models, however, like the dimer or multi-mer case 
\cite{flores,dunlap,bovier,hilke,evan,sanchez,hilke2}, this is not an 
essential condition.

The main conclusions we can derive from this study is that if we consider an Anderson model 
with every $d$'s site  disordered instead of each site, where $d$ is an integer and 
$d\geq2$, the model exhibits extended 
states at some critical energies. The exponents describing the strength of the divergence 
remain the same for the different energies, i.e., $\nu=2/3$. The delocalization properties of these diluted
random systems 
 can be understood in terms of correlations, as diluting the 
system is equivalent to introducing a long range periodic correlation in the disorder. Outside of the 
critical energies this dilute Anderson model has the same localization properties as the 
usual one.  When we equate the localization 
length with the size of the system, in order to estimate the number of states whose localization 
length exceeds the system size, we observe using (13) that this number is independent 
of the size, therefore  
in the infinite size  limit, these states shouldn't have any influence on the transport properties.
 But for small quasi-one dimensional systems like for example disordered 
superlattices of heterostructures or systems with very few impurities these effects do 
influence the transport properties. Finite temperatures can also reduce the 
effective system size and lead to changes in the transport properties.
The results presented above have important consequences on discretization 
procedures of disordered systems. Indeed, for a given number of disordered 
sites, the choice of the elementary lattice constant, drastically affects the localization 
properties of the system.

I would like to acknowledge C.P. Enz for helpfull discussions. This work 
was supported in part by the Swiss National Science Foundation.

\end{document}